\documentclass[%
 reprint,
 amsmath,amssymb,
 aps,
prb,
]{revtex4-2}

\usepackage[T1]{fontenc}

\usepackage{graphicx}
\usepackage{dcolumn}
\usepackage{bm}
\usepackage{booktabs} 
\usepackage{xcolor}
\usepackage{physics}   
\usepackage[justification=raggedright]{caption}
\usepackage[colorlinks=true, linkcolor=blue, citecolor=blue, urlcolor=blue]{hyperref}

\begin{document}

\title{High-throughput superconducting $T_{\mathrm{c}}$ predictions through density of states rescaling}

\author{Kieran Bozier}
    \email{kb697@cam.ac.uk}
\affiliation{Department of Materials Science and Metallurgy, University of Cambridge}
\author{Kang Wang}
 \affiliation{Department of Materials Science and Metallurgy, University of Cambridge}
\author{Bartomeu Monserrat}
 \affiliation{Department of Materials Science and Metallurgy, University of Cambridge}
\author{Chris J. Pickard}
 \email{cjp20@cam.ac.uk}
 \affiliation{Department of Materials Science and Metallurgy, University of Cambridge}
 \affiliation{Advanced Institute for Materials Research, Tohoku University, Sendai, 980-8577, Japan}

\date{\today}

\begin{abstract}

 First principles computational methods can predict the superconducting critical temperature $T_{\mathrm{c}}$ of conventional superconductors through the electron-phonon spectral function. Full convergence of this quantity requires Brillouin zone integration on very dense grids, presenting a bottleneck to high-throughput screening for high $T_{\mathrm{c}}$ systems. In this work, we show that an electron-phonon spectral function calculated at low cost on a coarse grid yields accurate $T_{\mathrm{c}}$ predictions, provided the function is rescaled to correct for the inaccurate value of the density of states at the Fermi energy on coarser grids. Compared to standard approaches, the method converges rapidly and improves the accuracy of predictions for systems with sharp features in the density of states. This approach can be directly integrated into existing materials screening workflows, enabling the rapid identification of promising candidates that might otherwise be overlooked.
\end{abstract}

\maketitle

\section{\label{sec:level1}Introduction}
The search for materials with higher superconducting transition temperatures ($T_{c}$) at ambient conditions is one of the great challenges in condensed matter physics. To date, despite decades of discovery, the majority of superconductors remain restricted to temperatures below liquid nitrogen ($77$\,K) or require the high-pressure environment of diamond anvil cells~\cite{boeri_2021_2022, pickard_superconducting_2020}. The complexity of this challenge has driven the development of sophisticated theoretical and computational tools to guide experimental efforts \cite{eliashberg_g_m_interactions_1960, scalapino_strong-coupling_1966, mcmillan_transition_1968, allen_transition_1975, oliveira_density-functional_1988}. A particularly powerful approach in the search for new conventional superconductors is first-principles calculations based on density functional theory (DFT) and Eliashberg theory \cite{margine_anisotropic_2013}. By calculating the electronic structure of a crystalline material, it is possible to calculate the phonon spectrum and evaluate the electron-phonon matrix elements, which describe the strength of the coupling between the electron and phonon subsystems~\cite{giustino_electron-phonon_2017, baroni_phonons_2001, monserrat_electronphonon_2018}. In conventional phonon-mediated BCS superconductors, the strength of this coupling determines the superconducting transition temperature~\cite{bardeen_theory_1957}. The advances in calculating electron-phonon properties from DFT have made it possible to predict $T_{\mathrm{c}}$ for conventional superconductors entirely from first principles~\cite{duan_ab_2019}. This has culminated in a string of successes in the field of high-pressure hydride superconductors, where in many cases the theoretical predictions have guided experimental discovery. Such systems include H$_3$S~\cite{drozdov_conventional_2015, duan_pressure-induced_2014}, CaH$_6$~\cite{hou_quantum_2023, wang_superconductive_2012}, and LaH$_{10}$~\cite{peng_hydrogen_2017, liu_potential_2017, errea_quantum_2020, somayazulu_evidence_2019, kruglov_superconductivity_2020,drozdov_superconductivity_2019}, all which display excellent agreement between the experimental and calculated $T_{\mathrm{c}}$.

These first principles calculations are now commonly used alongside methods for crystal structure prediction~\cite{oganov_structure_2019}, such as \textit{ab initio} random structure searching~\cite{pickard_structures_2009, pickard_ab_2011, dolui_feasible_2024}, minima hopping~\cite{goedecker_minima_2004, sanna_prediction_2024}, generative models~\cite{zeni_generative_2025}, particle swarms~\cite{wang_calypso_2012,liu_potential_2017} and genetic algorithms~\cite{glass_uspexevolutionary_2006, ishikawa_materials_2019, lonie_xtalopt_2011} to explore increasingly large regions of composition and pressure space for promising superconducting materials~\cite{dolui_feasible_2024, sanna_prediction_2024, hutcheon_predicting_2020, shipley_stability_2020, zurek_hydrides_2017}. Many of these methods are now leveraging machine-learning to accelerate these searches and explore more complex candidates~\cite{pickard_ephemeral_2022,salzbrenner_developments_2023,merchant_scaling_2023,wang_concurrent_2024, podryabinkin_accelerating_2019}. However, high-throughput screening comes with some challenges. When screening through many hundreds or thousands of structures, the computational cost of finding the property of interest for each structure must be as low as possible. However, a fully converged first principles calculation of the phonon and electron-phonon properties of a material requires very fine electronic $\mathbf{k}$ and phonon $\mathbf{q}$ grids, which is prohibitively expensive~\cite{giustino_electron-phonon_2007}. Using coarser electron and phonon grids keeps the computational cost low, but can compromise the accuracy. Therefore, a balance must be achieved, between screening as many candidates as possible, and ensuring the predicted $T_{\mathrm{c}}$ from the coarse grids remains close to the converged value.

The reality is that many superconducting systems are poorly described by these coarse grid calculations. One of the most critical ingredients for a high $T_{\mathrm{c}}$ is a large density of states (DOS) at the Fermi energy, denoted $N_{\mathrm{F}}$~\cite{boeri_2021_2022}. This is because the presence of more states at the Fermi energy allows more electrons to participate in the BCS pairing interaction~\cite{cooper_bound_1956, schrieffer_theory_1999}. Other factors are also important, such as high-frequency phonons leading to a large $\omega_{\text{log}}$ in the McMillan and Allen-Dynes equations, but these tend to be subject to fundamental limits set by physical constants~\cite{trachenko_upper_2025} and thus the ability to optimise these is more limited. Consequently, most promising systems have a high density of states at the Fermi energy. This commonly occurs in proximity to a sharp peak in the density of states - although these systems do often host competing instabilities such as magnetic ordering and charge density waves. Sharp peaks in the DOS are rarely well captured by coarse electronic grids, because in most superconductivity calculations the density of states is modeled by applying Gaussian smearings with a fixed width $\sigma$ to the energy eigenvalues from the electronic grid. This method can produce unrepresentative values of the density of states when the electronic grid is too coarse \cite{pickard_extrapolative_1999, toriyama_how_2022}, leading to a direct under- or overestimation in $T_{\mathrm{c}}$. Examples of high $T_{\mathrm{c}}$ systems with a sharp peak near the Fermi energy are H$_3$S~\cite{quan_van_2016} and the proposed candidate Mg$_2$IrH$_6$~\cite{dolui_feasible_2024, sanna_prediction_2024}, and both result in substantially underestimated critical temperatures when coarse grids are used. In the context of high-throughput screening, there is therefore a risk that an interesting candidate with an anomalously high density of states would be filtered out of the search at an early stage.

The importance of the density of states in the convergence of $T_{\mathrm{c}}$ was explored by Koretsune and Arita in Ref.~\cite{koretsune_efficient_2017}. They argue that many quantities important for superconductivity can be written in the form of the density of states multiplied by a weighted average. For many systems, they found that this average term shows only a weak dependence on the electronic $\mathbf{k}$ grid and smearing $\sigma$, demonstrating that it is often the convergence of $N_{\mathrm{F}}$ that is responsible for the very fine grids required in $T_{c}$ calculations. In a later work~\cite{morice_weak_2017}, they apply this method to the BiS$_2$ system to obtain an electron-phonon coupling strength $\lambda$ and $T_{\mathrm{c}}$ that were almost independent of smearing, by multiplying the electron-phonon spectral function $\alpha^2F(\omega)$ by a correction factor. In this work, we directly build on these ideas, namely how describing the density of states correctly is transformative in accelerating convergence on coarse grids - with a particular focus on its importance in high-throughput calculations. By introducing a correction factor to rescale the relevant superconducting functions, as was done in Ref.~\cite{morice_weak_2017}, we have been able to accelerate convergence with respect to the Gaussian smearing width and grid sizes. Importantly, we have found the method is most effective for systems with sharp features in the density of states, which are among the most promising systems for high-temperature superconductivity.

\section{Theory}
The central quantity in the calculation of the transition temperature is the Eliashberg electron-phonon spectral function $\alpha^2F(\mathbf{k}, \mathbf{k'}, \omega)$. This represents the average of the electron-phonon matrix elements over the allowed momentum transfers~\cite{scalapino_strong-coupling_1966}. The function, which is proportional to the phonon density of states weighted by the electron-phonon matrix elements $g_{nm\nu}(\mathbf{k},\mathbf{k'})$, is given by~\cite{pellegrini_ab_2024, allen_theory_1983}
\begin{equation}
\begin{split}
\alpha^{2}F_{nm}(\mathbf{k}, \mathbf{k'}, \omega) = N_{\mathrm{F}} \sum_{\nu}& \left| g_{nm\nu}(\mathbf{k},\mathbf{k'})\right|^2 \\ 
& \times\delta(\omega - \omega_{\mathbf{k} - \mathbf{k'}, \nu}) .
\end{split}
\end{equation}
It is linearly dependent on $N_{\mathrm{F}}$, the density of electronic states at the Fermi energy. Here, $\mathbf{k}$ and $\mathbf{k'}$ are the electronic momenta, $n, m$ are the electronic band indices, and $g_{nm\nu}(\mathbf{k},\mathbf{k'})$ are the electron-phonon matrix elements for the electronic states $\ket{\mathbf{k}, n}$ and $\ket{\mathbf{k'}m}$ coupled by a phonon of mode $\nu$, momentum $\mathbf{q} = \mathbf{k}-\mathbf{k'}$ and frequency $\omega_{\mathbf{k}-\mathbf{k'}, \nu}$.

In the isotropic approximation, the average value of $\alpha^2F(\omega)$ for all $\ket{\mathbf{k}, n}$ and $\ket{\mathbf{k'}m}$ on the Fermi surface is used. This can be obtained by inserting dimensionless weighting functions $W_{\mathbf{k}n} = \frac{\delta(\epsilon_{\mathbf{k}n})}{N_{\mathrm{F}}}$, where $\epsilon_{\mathbf{k}n}$ is the energy with respect to the Fermi energy, and performing a summation over all momenta $\mathbf{k},\mathbf{k'}$ and bands $n, m$. This leads to the more commonly seen expression~\cite{pellegrini_ab_2024}
\begin{equation}
    \begin{split}
    \alpha^2F(\omega) = \frac{1}{N_{\mathrm{F}}}& \sum_{\mathbf{k}, \mathbf{k'},n,m} \sum_{\nu}  \;\left| g_{nm\nu}(\mathbf{k},\mathbf{k'})\right|^2  \\
    &\quad \times \,\delta(\epsilon_{\mathbf{k}n}) \delta(\epsilon_{\mathbf{k'}m}) \delta(\omega - \omega_{\mathbf{k} - \mathbf{k'}, \nu}).
    \end{split}
    \label{eqn:alpha2F-long}
\end{equation}

The function is still directly proportional to the density of electronic states at the Fermi level because each summation over $\mathbf{k}, n$ of the delta functions $\delta(\epsilon_{\mathbf{k}n})$ yields a factor of $N_{\mathrm{F}}$. The other quantity commonly used in calculations of $T_{\mathrm{c}}$ is the isotropic electron-phonon coupling strength $\lambda$, given by
\begin{equation}
    \lambda = \int d\omega \frac{2\alpha^2F(\omega)}{\omega},
    \label{eqn: lambda}
\end{equation}
where again, this quantity is linearly proportional to the density of states at the Fermi energy. Hence, in calculations of these quantities, the electronic states used in the Brillouin zone integrals must be able to reproduce the correct density of states of the material. The $T_{\mathrm{c}}$ can then be found using the coupling strength, most commonly using one of the McMillan equation~\cite{mcmillan_transition_1968}, the Allen-Dynes equation~\cite{allen_transition_1975}, a machine-learned model~\cite{hutcheon_predicting_2020, xie_machine_2022}, or by directly solving the Eliashberg equations~\cite{margine_anisotropic_2013}.

To compute these functions on finite electronic $\mathbf{k}$ grids, the Dirac delta functions are replaced with Gaussians of finite width. The particular expression used in the \textsc{Quantum Espresso} code~\cite{giannozzi_quantum_2009} is given by
\begin{equation}
	\delta(\epsilon_{\mathbf{k}n}) \rightarrow \frac{1}{\sigma \sqrt{\pi}} \exp( \frac{\epsilon_{\mathbf{k}n}^2}{\sigma^2}) .
\end{equation}
Selecting the smearing $\sigma$ requires a convergence test. The Dirac-delta function corresponds to infinitesimal smearing, but on finite grids very small smearings lead to disconnected points. By comparison, overly broad smearings may converge faster but to an incorrect value, as points far from the Fermi energy are included in the summation~\cite{pellegrini_ab_2024}. Several methods are commonly used to select the smearing: (i) plotting $T_{\mathrm{c}}$ against smearing for different electronic grids, and the optimal value chosen as the smallest smearing where the curves do not diverge~\cite{wierzbowska_origins_2006,shipley_stability_2020}; (ii) for high-throughput searching, plot just one of these curves and take the smallest smearing where $T_{\mathrm{c}}$ does not rapidly diverge from the large smearing values; (iii) choose the Gaussian smearing value that most closely reproduces the value of $N_{\mathrm{F}}$ found from a very well converged density of states calculation \cite{koretsune_efficient_2017}, prioritizing the importance of an accurate density of states. All rely on $T_{\mathrm{c}}$ varying only slowly as a function of smearing $\sigma$, so methods that can reduce this dependence are advantageous.

This procedure of replacing the Dirac-delta functions with Gaussians is also commonly used in calculations of the density of states \cite{toriyama_how_2022}. However, it is not a particularly effective method, often requiring extremely dense electronic $\mathbf{k}$ grids to produce a converged plot \cite{pickard_extrapolative_1999}. Quantities that depend directly on the density of electronic states at the Fermi energy, such as $\alpha^2F(\omega)$, will thus often require very fine electronic grids if calculated with the Gaussian approach. The main drawback of the method is that all $\mathbf{k}$-points have the same smearing applied, meaning sharp features in the density of states corresponding to flat bands tend to become smeared out. Instead, more effective approaches are used such as adaptive smearing~\cite{morris_optados_2014, nicholls_optados_2012}, extrapolative methods~\cite{pickard_extrapolative_1999,pickard_second-order_2000} or tetrahedra methods~\cite{blochl_improved_1994, kawamura_improved_2014}. By taking advantage of these more effective ways of calculating $N_{\mathrm{F}}$, it should be possible to more rapidly converge $\alpha^2F(\omega)$ and $\lambda$.

\section{Method and results}
\subsection{Calculation details}
Calculations of the electron, phonon, and electron-phonon properties were performed with the \textsc{Quantum Espresso} code~\cite{giannozzi_quantum_2009} using the pseudo-dojo \textsc{PBE oncvpsp} pseudopotentials~\cite{hamann_optimized_2013, van_setten_pseudodojo_2018}. The kinetic energy cutoff was determined by converging the total energy to $< 1$\,meV per atom, requiring cutoffs of around 100\,Ry. Except where otherwise specified, the phonons were calculated on a coarse grid with a $\mathbf{q}$ point spacing of $(0.06\times2\pi)$\,\AA$^{-1}$ and a coarse $\mathbf{k}$ point spacing of $(0.02\times2\pi)$\,\AA$^{-1}$, which was then interpolated onto a fine electron grid with spacing $(0.01\times2\pi)$\,\AA$^{-1}$ to calculate the electron-phonon matrix elements $g_{nm\nu}(\mathbf{k},\mathbf{k'})$ and coupling strength $\lambda$. The $T_{\mathrm{c}}$ was calculated by solving the isotropic Eliashberg equations in the constant DOS approximation, with a fixed Coulomb pseudopotential $\mu^* = 0.125$. In principle, the value of $\mu^*$ should be calculated from the screened Coulomb interaction~\cite{pellegrini_ab_2023, pellegrini_ab_2024, kogler_isome_2025}, but this is computationally challenging in high-throughput searches and therefore not explored here.

\subsection{Rescaling by the density of states}
To accelerate convergence, we perform an accurate calculation of the density of states at the Fermi energy using the linear-interpolated tetrahedra method, and use this to replace the value obtained by the Gaussian smearing. This can be performed as a post-processing step \cite{morice_weak_2017}, by multiplying $\alpha^2F(\omega)$ by a simple prefactor
\begin{equation}
    \widetilde{\alpha^2F}(\omega, \sigma) =  \frac{\widetilde{N_{\mathrm{F}}}}{N_{\mathrm{F}}(\sigma)} \; \alpha^2F(\omega, \sigma),
    \label{eqn: scaling_factor}
\end{equation}
where $\widetilde{N_{\mathrm{F}}}$ is a high quality density of states calculated with the tetrahedra method, $N_{\mathrm{F}}(\sigma)$ is the density of states calculated with the Gaussian smearing used in the $T_{\mathrm{c}}$ calculation, and $\widetilde{\alpha^2F}$ is the rescaled Eliashberg spectral function. In the case of a very well-converged smearing DOS, this prefactor will be unity and this procedure has no effect. However, in high-throughput calculations the $T_{c}$ and Gaussian DOS will typically not be fully converged due to the cost of using a very fine electronic grid. This prefactor therefore aims to correct the error arising from a poorly described DOS. As the cost of an accurate DOS calculation is usually very small compared to the phonon calculation, and is typically already performed as part of a high-throughput screening to remove non-metallic systems, this step can easily be incorporated into existing workflows. The rescaling method is expected to show less variation in $T_{\mathrm{c}}$ with respect to smearing $\sigma$, since the smearing dependence of $N_{\mathrm{F}}$ has been removed. Nevertheless, some variation will still remain, because smearing controls which $\mathbf{k},\mathbf{k'}$ values in equation~\ref{eqn:alpha2F-long} are included.

\subsection{Limitations of Gaussian smearing}

\begin{figure}[th!]
    \centering
    \includegraphics[width=\linewidth]{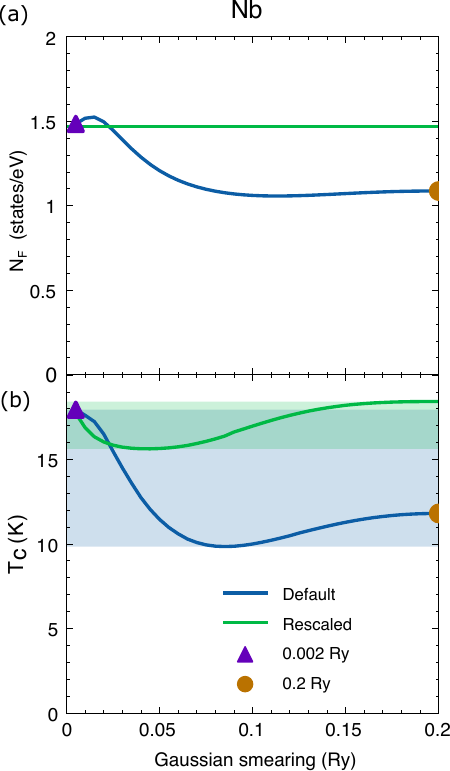}
    \caption{The $T_{\mathrm{c}}$ is strongly dependent on the density of states at the Fermi level. In (a), the density of states at the Fermi energy $N_{\mathrm{F}}$ and in (b), the $T_{\mathrm{c}}$ are plotted as a function of the smearing width of the Gaussian broadening. The blue line is for the standard Gaussian approach, while the green line has had the rescaling factor in equation \ref{eqn: scaling_factor} applied. The purple and orange points indicate the smearings used for the density of states in Figure \ref{fig:type5-DOS}.}
    \label{fig:type5-NfTc}
\end{figure}

\begin{figure}
    \centering    \includegraphics[width=0.95\linewidth]{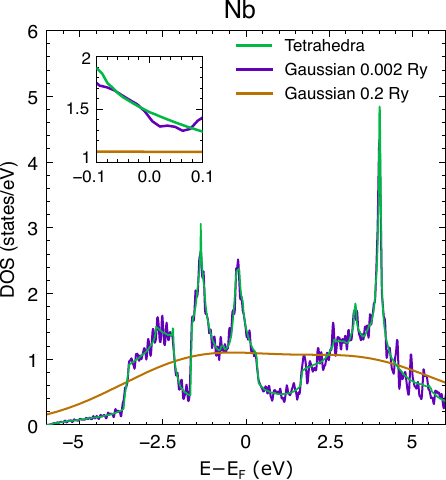}
    \caption{The density of states of Nb is plotted for a converged tetrahedra calculation (green), and the Gaussian methods with an overly narrow (purple) and an overly broad (orange) smearing width.}
    \label{fig:type5-DOS}
\end{figure}

Figures \ref{fig:type5-NfTc} and \ref{fig:type5-DOS} demonstrate how an inappropriate smearing with the Gaussian approach can lead to an incorrect density of states, and consequently ${T}_{c}$ for niobium. In subplots (a) and (b) in figure \ref{fig:type5-NfTc}, the density of states at the Fermi level and corresponding $T_{c}$ are plotted as functions of the Gaussian smearing for both the standard Gaussian method (shown in blue) and with the rescaling factor in equation \ref{eqn: scaling_factor} applied (shown in green). It is clear that the rescaled method displays less variation in $T_{\mathrm{c}}$ as the smearing is varied. In this example, the large smearing of $0.2$\,Ry leads to any sharp peaks in the DOS being smoothed out, as can be observed in the orange line in figure~\ref{fig:type5-DOS}. As the Fermi energy in Nb lies on the downward slope of a sharp peak in the DOS, this leads to a substantial underestimation of $N_{\mathrm{F}}$ as seen in the inset, and correspondingly a lower value of $T_{\mathrm{c}}$. If instead the Fermi energy is close to a valley, the overly-smooth density of states would overestimate $N_{\mathrm{F}}$ and $T_{\mathrm{c}}$. In comparison, if the smearing used is too small, as illustrated by the $0.002$\,Ry data, it can introduce spurious oscillations into smooth regions of the density of states. This is seen in the purple line of figure \ref{fig:type5-DOS}, particularly around 5\,eV, and again can result in over- or underestimation of $N_{\mathrm{F}}$ and $T_{\mathrm{c}}$. These oscillations arise due to insufficient sampling of the Brillouin zone, and can result in very large variations in $T_{\mathrm{c}}$ since as the smearing is reduced the density of states begins to resemble a set of disjointed Dirac delta functions. For this particular system and sampling, the optimal choice of smearing parameter to reproduce the converged tetrahedra density of states is $0.025$\,Ry. However, it is important to note that this optimal value depends both on the sampling density, and the curvature of the DOS at the Fermi level, with denser grids and sharper features necessitating smaller smearings.

It should be briefly noted that the obtained converged value of $T_\mathrm{c}$ in niobium for a $\mu^*=0.125$ is 16\,K, in agreement with other computational work~\cite{pellegrini_ab_2024}. However, this is notably larger than the experimental value of $9.2$\,K~\cite{finnemore_superconducting_1966}, which is attributed to the need for a larger $\mu^*$ value in this system~\cite{pellegrini_ab_2024,kogler_isome_2025}. Nonetheless, the obtained converged value of the electron-phonon coupling strength with rescaling is $\lambda = 1.4$, which is in good agreement with experimental data~\cite{allen_empirical_1987}.

\begin{figure*}[t!]
    \centering
    \includegraphics[width=\linewidth]{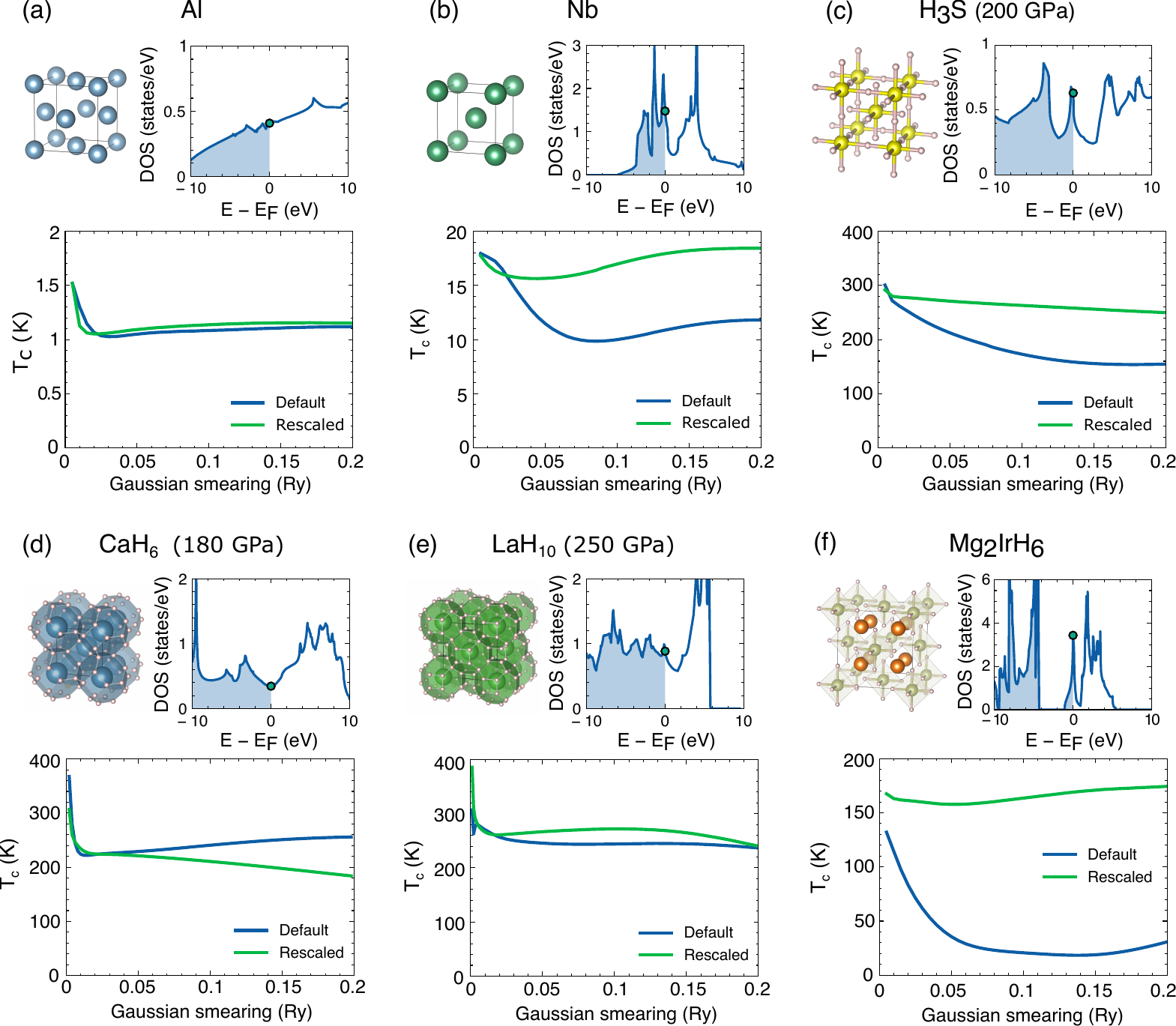
    }
    \captionsetup{width=\textwidth}
   
    \caption{The shape of the density of states at the Fermi level determines the impact of the rescaling, shown for (a) Al, (b) Nb, (c) H$_3$S, (d) CaH$_6$, (e) LaH$_{10}$ and (f) Mg$_2$IrH$_6$ . Each panel shows the structure visualized with VESTA~\cite{momma_vesta_2011}, the density of states of the primitive cell, and the Eliashberg $T_{c}$ as a function of the Gaussian smearing for the default Gaussian method and the rescaled approach. The sharper the peaks near the Fermi energy, the greater the impact of rescaling, with Mg$_2$IrH$_6$ being the system where rescaling has the largest impact.}
    \label{fig:type4}
\end{figure*}

\subsection{Impact of rescaling}
The degree to which the rescaling impacts the $T_{\mathrm{c}}$ varies between systems, and is dependent on the shape of the density of states. Figure \ref{fig:type4} shows a set of panels for six superconducting systems. Each panel shows the conventional cell visualized with VESTA~\cite{momma_vesta_2011}, a high quality density of states calculated with the tetrahedra method on electronic grids with $60^3$ points, and plots of the $T_{\mathrm{c}}$ as a function of Gaussian smearing for both the default and rescaled methods. Starting from the top left, in panel (a) we show aluminium, which is a system that is already well described by the standard Gaussian approach. The smooth density of states means that $N_{\mathrm{F}}$ is already well described even for broad smearings, resulting in a well converged value of $T_{c}$ with respect to smearing for both methods. Niobium is presented in panel (b) where the Fermi energy lies at a downward slope of a sharp peak. Consequently, the rescaling causes an increase in $T_{\mathrm{c}}$ at broad smearings, and the green rescaled curve appears to be flatter, indicating better convergence. Panel (c) shows H$_3$S, where like niobium, the Fermi energy lies near a peak and results in an up-shift in $T_{\mathrm{c}}$ after rescaling. Additionally, the convergence plot shows the rescaled $T_{c}$ is significantly less dependent on the smearing - indicating that for H$_3$S, the variation in the electron-phonon matrix elements is small and so convergence is dominated by how well the density of states is described. In (d), we see CaH$_6$, where the Fermi energy lies in a shallow valley of the density of states, and so a resulting minor downshift in $T_{c}$ is seen at large smearings. Both methods present well converged results and are in good agreement with each other below 0.05\,Ry. Panel (e) shows LaH$_{10}$, which is one of only a few systems where the rescaling arguably hinders convergence. This is discussed in section~\ref{sec:level4}, and is hypothesized to be due to a cancellation of errors in the default Gaussian method. Nonetheless, the high $T_{\mathrm{c}}$ for both approaches means this structure would not be missed in high-throughput searching. Lastly, we present Mg$_2$IrH$_6$ in (f), where the rescaling is at its most transformative. Here, the Fermi energy lies on a very sharp peak, which all but the smallest smearings will drastically over-smear. This results in a substantial underestimation of $T_{\mathrm{c}}$ if the standard method is used. In contrast, the rescaling easily picks out this system as having a very high $T_{c}$, with a value of about $160$\,K for this choice of $\mu^*$, in very good agreement with previous studies that converged the $T_{\mathrm{c}}$ on exceptionally fine grids~\cite{dolui_feasible_2024}. Importantly in the context of structure searching and high-throughput screening, if the default methods were used with a smearing that is too broad, this structure would likely be prematurely removed from the list of promising candidates.


\subsection{Rescaling for high-throughput}
\begin{figure}[t]
    \centering
    \includegraphics[width=\linewidth]{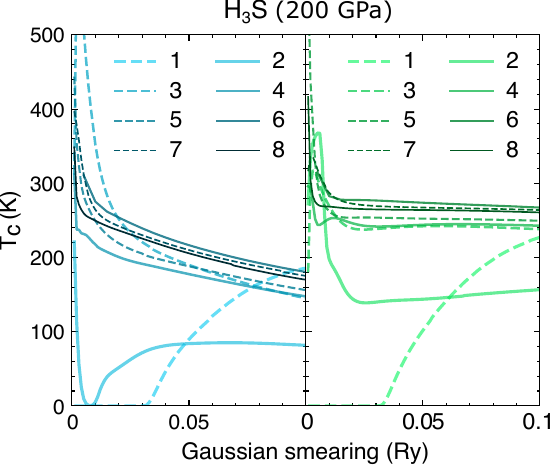}
    \caption{Comparison between convergence of $\text{H}_3\text{S}$ with respect to the phonon $q\times q\times q$ grid using the default method (left) and with the rescaled approach (right). The electronic grid is set to four times the density of the phonon grid. The rescaled method displays a noticeable faster convergence, illustrated by the flatter curves.}
    \label{fig:type6-H3S}
\end{figure}

\begin{figure}[t]
    \centering
    \includegraphics[width=\linewidth]{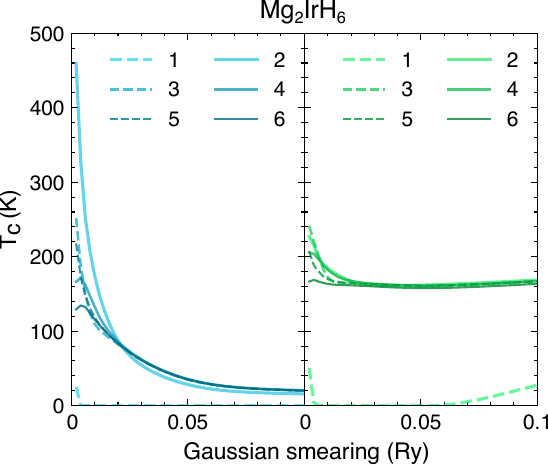}
    \caption{Comparison between convergence of $\text{Mg}_2\text{IrH}_6$ with respect to the phonon $q\times q\times q$ grid using the default method (left) and with the rescaled approach (right). Here the rescaling achieves the converged result of 160\,K~\protect{\cite{dolui_feasible_2024}} even for a $2^3$ grid.}
    \label{fig:type6-Mg2IrH6}
\end{figure}

\begin{figure}
    \centering
    \includegraphics[width=\linewidth]{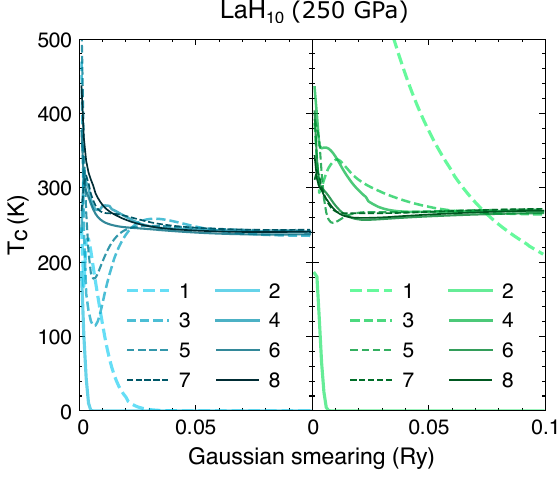}
    \caption{Comparison between convergence of $\text{LaH}_{10}$ with respect to the phonon $q\times q\times q$ grid using the default method (left) and with the rescaled approach (right). For this system, the rescaling does not enhance convergence.}
    \label{fig:type6-LaH10}
\end{figure}

There is an additional benefit to the rescaling when used for high-throughput searches; not only does it pick out systems with high values of $N_{\mathrm{F}}$, but it often accelerates the convergence with respect to the electronic grid size. This is because in many systems, the change in $N_{\mathrm{F}}$ as the grid is varied is the dominant contribution to the convergence of $T_{c}$. As stated previously, using fine grids for calculations of $\lambda$ and $T_{\mathrm{c}}$ is prohibitively expensive if applied to several hundreds or thousands of candidates. As such, a rapid convergence of $T_{c}$ with respect to the phonon $\mathbf{q}$ grid and the electron $\mathbf{k}$ grid is highly desirable. To this end, we compare the $T_{\mathrm{c}}$ for different $\mathbf{q}$ (and corresponding $\mathbf{k}$ grids) for both the standard and rescaling methods. Figure \ref{fig:type6-H3S} shows the results for $\text{H}_3\text{S}$. The left plot with the blue curves shows the $T_{\mathrm{c}}$ against smearing $\sigma$ for the standard Gaussian smearing method, with a range of grid sizes ranging from a $1^3$ to an $8^3$ phonon grid. The coarse electronic grid used for the phonon calculation is set to twice the phonon grid density, and the fine electronic grid for the matrix element interpolation is set to four times the density. The right plot with the green curves shows the same results, but for the rescaled method. As expected, in both methods, insufficient sampling (such as a $\Gamma$ point or a $2^3$ grid) leads to a wide range of values of $T_{\mathrm{c}}$. However, for a $3^3$ grid and finer, the rescaled method produces a smaller variation in $T_{\mathrm{c}}$ as a function of the smearing, and the converged value of $250\,\pm\,10$\,K agrees reasonably well with the literature~\cite{errea_high-pressure_2015,duan_ab_2019}. This improvement when using the rescaling method is indicative that for the $\text{H}_3\text{S}$ system, it is the density of states that is the limiting factor in achieving convergence. The impact is more stark for Mg$_2$IrH$_6$ in figure \ref{fig:type6-Mg2IrH6}, with even $2^3$ grids reproducing the converged value of 160\,K~\cite{dolui_feasible_2024}, and $T_{\mathrm{c}}$ is almost independent of the smearing parameter. However, an improvement is not guaranteed in all cases. In figure \ref{fig:type6-LaH10} we present the same convergence tests but for the $\text{LaH}_{10}$ system. For this system, both the standard and rescaled methods produce very similar curves, with a converged result of 250\,K in agreement with the literature~\cite{peng_hydrogen_2017}. This indicates that in this case the density of states is quite smooth and already well resolved even for large smearings, and so convergence is dominated instead by variations in the electron-phonon matrix elements $|g_{nm\nu}(\mathbf{k},\mathbf{k'})|^2$, which are unaffected by the rescaling procedure.

\section{\label{sec:level4}Discussion}

As shown earlier, the success and impact of the rescaled DOS method varies depending on the system. The method is most accurate in systems where the convergence of $\lambda$, and correspondingly $T_{\mathrm{c}}$, is dominated by the calculated value of $N_{\mathrm{F}}$. This corresponds to systems where the electron-phonon matrix elements $g_{nm\nu}(\mathbf{k},\mathbf{k'})$ exhibit only weak momentum and energy dependence near the Fermi surface. In the idealized limit of constant matrix elements $g_{nm\nu}(\mathbf{k},\mathbf{k'}) = g$ and neglecting the momentum and band dependence of the phonon frequencies $\omega_{\mathbf{q},\nu} = \omega$, the expression in equation \ref{eqn: lambda} simplifies to
\begin{equation}
    \lambda(\sigma) =  \frac{2  |g|^2}{\omega} \,N_{\mathrm{F}}(\sigma).
\end{equation}
The only parameter dependent on smearing is $N_{\mathrm{F}}$, so the rescaled DOS method achieves its highest accuracy, since it eliminates the error introduced by a poorly described density of states. Importantly, as discussed previously, many high-throughput workflows will calculate an accurate DOS to screen out non-metallic systems, so this more accurate value of $N_{\mathrm{F}}$ can be used for rescaling at no additional cost.

The impact of the rescaling is determined by the shape of the density of states near the Fermi energy, and by how well this region is described by Gaussian smearing. In principle, any density of states can be replicated with sufficiently fine electronic sampling using the standard Gaussian smearing approach~\cite{morris_optados_2014}. However, in practice, the computational constraints of high-throughput searching mean that the grid used is almost always too coarse. This causes inaccuracies in the value of $N_{\mathrm{F}}$. Systems with smooth, broad features in their density of states, such as aluminium in figure \ref{fig:type4}(a), will not benefit from the rescaling, as these are already well described by broad Gaussian smearings. However, if there are sharp features, such as narrow peaks in the vicinity of the Fermi energy, then rescaling can be transformative. In the special case of $\text{Mg}_2\text{IrH}_6$, where the Fermi energy lies on a peak, the rescaling allows us to reproduce the converged result for this level of theory of $160$\,K~\cite{dolui_feasible_2024} at a fraction of the cost (a speed-up factor of 120 between the $2^3$ and $5^3$ grids). More importantly, as seen in figure \ref{fig:type4}(f) and figure \ref{fig:type6-Mg2IrH6}, without the rescaling factor the obtained $T_{c}$ is far below the converged value - even for a reasonably dense grid. It is only with extremely dense grids and narrow smearings, such as those obtained with the electron-phonon Wannier code {\sc epw}~\cite{lee_electronphonon_2023}, that the high $T_{\mathrm{c}}$ emerges when the standard method is used. Such systems with very sharp peaks at the Fermi energy are of particular importance in searches for high $T_{c}$ superconductors, as the high density of states leads to large electron-phonon coupling strengths~\cite{boeri_2021_2022}. However, the standard Gaussian smearing approach used in current workflows can significantly underestimate $N_{\mathrm{F}}$ and consequently $T_{c}$, particularly on coarse grids, meaning these promising candidates may be overlooked in the search. 

Rescaling can also be beneficial for accelerating convergence with respect to grid sizes. Although the coupling strength $\lambda$ on a coarse grid will generally deviate from the fine grid value - due to some $\mathbf{q}$-points being omitted from the weighted sum - in the test cases investigated the difference was within 20\,\%. This led to estimates of $T_{\mathrm{c}}$ in figures \ref{fig:type6-H3S}, \ref{fig:type6-Mg2IrH6} and \ref{fig:type6-LaH10} that were largely in agreement with the literature, even on coarse grids. The value of rescaling here is twofold. First, the removal of the smearing dependency of the density of states typically leads to a better convergence of $\lambda$ with respect to smearing, making it possible to obtain a consistent $\lambda(\sigma)$ even on coarse grids. Second, the rescaling removes the need to use very fine electronic grids to properly converge the $N_{\mathrm{F}}$ value, allowing the use of coarser $\mathbf{k}$ grids. Often a fixed ratio is used between the $\mathbf{k}$ and $\mathbf{q}$ grids, so this also means coarser $\mathbf{q}$ grids can be used, provided the phonon properties converge relatively quickly.

It is important to note some of the limitations of the method. In certain cases, such as LaH$_{10}$ in figure \ref{fig:type4} (e), the standard Gaussian method already produces convergence in $T_{c}$ with respect to smearing width for relatively large (greater than $0.05$\,Ry) broadening. The inclusion of the rescaling factor slightly worsens the result, although the $T_{\mathrm{c}}$ is still well converged. We suggest a partial cancellation of errors may be responsible. The two sources of error are the value of the matrix elements, and their weighting given by the smeared delta functions. For the former, the finite $\mathbf{k}$-point grid means values of the matrix elements $|g_{nm\nu}{(\mathbf{k},\mathbf{k'})}|$ are not directly sampled on the Fermi surface, but at points some distance away. If the value of $|g_{nm\nu}{(\mathbf{k},\mathbf{k'})}|$ on the sampled points does not reflect the true value on the Fermi surface then a systematic error is introduced. In particular, at larger broadening, points increasingly far from the Fermi energy with potentially less representative values of $|g_{nm\nu}{(\mathbf{k},\mathbf{k'})}|$ are included~\cite{pellegrini_ab_2024}. Concurrently, the actual sampling and smearing of the electronic points introduces its own error, namely that it may not accurately reflect the true density of states at the Fermi energy. This manifests as an incorrect weighting. When the rescaling factor is applied, it will correct for the second error, but not for the first. As such, in some cases it is possible that there may be a degree of error cancellation in the standard Gaussian method, which is then not present in the rescaled method. For this reason, we recommend plotting both the standard Gaussian and rescaled methods when investigating convergence, as comparing the two gives useful insight into which factor dominates convergence.

A further limitation is that sharp peaks in the density of states at the Fermi energy can cause an overestimation in $T_{\mathrm{c}}$ compared to the experimental value if rescaling is used. This is not as much a flaw with the rescaling as with the constant-DOS Eliashberg equations. These equations assume a flat DOS around the Fermi energy with a value of $N_{\mathrm{F}}$, which the rescaling may set too high \cite{sano_effect_2016}. In contrast, the standard Gaussian broadened method will typically underestimate $T_{\mathrm{c}}$ in these systems as the peak in the density of states is entirely smeared out. The solution is to use the more expensive variable-DOS Eliashberg equations~\cite{kogler_isome_2025, lucrezi_full-bandwidth_2024}, which are sensitive to the shape of the density of states around the Fermi energy. Testing with a variable-DOS method~\cite{kogler_isome_2025} tends to yield a $T_{\mathrm{c}}$ between these two extremes. As promising candidates from a high-throughput search must always be subjected to further analysis at higher levels of theory and tighter convergence before conclusions can be made, our view is it is better to overestimate these systems with pronounced peaks in the DOS at an early stage to prevent them from being prematurely screened out.
\\
\section{Conclusion}
In summary, rescaling the Eliashberg spectral function $\alpha^{2}F(\omega)$ by a factor to correct the density of states is a cheap and physically-motivated approach to improving the calculated value of $\lambda$ and $T_{\mathrm{c}}$. The method can enhance numerical convergence with respect to smearing parameters, and can correct for the error in $T_{\mathrm{c}}$ introduced by a poorly described density of states. The improvement is largest for systems with sharp features in the density of states at the Fermi energy, which is the case for many high $T_{\mathrm{c}}$ systems. These improved predictions, even at coarse grid sizes, allow better estimation of the $T_{\mathrm{c}}$ and ensure promising candidates are not screened out during high-throughput searching.

\begin{acknowledgments}
We thank P.I.C. Cooke and M. Caussé for many helpful discussions. K.B. gratefully acknowledges funding from an ESPRC DTP studentship in the Department of Materials Science and Metallurgy. 
K.W. and B.M. are supported by a UKRI Future Leaders Fellowship [MR/V023926/1]. B.M. also acknowledges support from the Gianna Angelopoulos Programme for Science, Technology, and Innovation.
We are grateful for computational support from the UK national high performance computing service, ARCHER2, for which access was obtained via the UKCP consortium and funded by EPSRC grant ref EP/X035891/1.\\
\\

\end{acknowledgments}

\bibliography{apssamp}

\end{document}